\documentclass{elsart}
\journal{Physica D}
\newcommand{\taq}{{\tilde{A}}_q}
\newcommand{\tbq}{{\tilde{B}}_q}
\newcommand{\bPsi}{{\psi^*}}
\newcommand{\hsi}{\hat{\psi}}
\newcommand{\bhsi}{\hat{\psi}^*}
\newcommand{\hso}{\hat{\psi}_o}
\newcommand{\bhso}{\hat{\psi}^*_o}
\newcommand{\bhsos}{\hat{\psi}^{*2}_o}
\newcommand{\dsi}{\delta\hat{\psi}}
\newcommand{\bdsi}{\delta{\psi^*}}
\newcommand{\beq}{\begin{equation}}
\newcommand{\eeq}{\end{equation}}
\newcommand{\bea}{\begin{eqnarray}}
\newcommand{\eea}{\end{eqnarray}}
\newcommand{\spin}[1]{{\bf \hat{#1}}}
\usepackage{amssymb}
\usepackage{graphicx}
\begin{document}
\begin{frontmatter}
\title{Equatorial  and related non-equilibrium states in magnetization 
dynamics of ferromagnets: Generalization of Suhl's spin-wave instabilities}

\author[tokyo]{C.~Kosaka},
\author[tokyo]{K.~Nakamura},
\author[bard]{S.~Murugesh},
\author[bard]{M.~Lakshmanan\corauthref{auth}}
\ead{lakshman@cnld.bdu.ac.in}
\corauth[auth]{Corresponding author.}

\address[tokyo]{Department of Applied Physics, Osaka City University, Osaka 
558-8585,
Japan}

\address[bard]{Centre for Nonlinear Dynamics, Department of Physics, 
Bharathidasan University, Tiruchirappalli 620024, India}

\begin{abstract}
We investigate the nonlinear dynamics underlying the evolution of a 2-D 
  nanoscale ferromagnetic film with uniaxial anisotropy in the presence of
perpendicular pumping.
Considering the associated Landau-Lifshitz spin evolution equation with
Gilbert damping   together with Maxwell equation for the demagnetization field,
we study the dynamics in terms of the stereographic
variable. We identify several new fixed points for suitable choice of external field
  in a rotating frame of reference. 
  In particular, we identify explicit equatorial and related 
fixed points of the spin vector in the plane transverse to the anisotropy axis 
when the pumping frequency coincides with the amplitude of the static parallel 
field. We then study the linear stability of these novel fixed points under 
homogeneous and spin wave perturbations and obtain a generalized Suhl's 
instability criterion, giving the condition for exponential growth of P-modes 
under spin wave perturbations.  Two parameter phase diagrams (in terms of 
amplitudes of static parallel and oscillatory perpendicular magnetic fields) 
for stability are obtained, which differ qualitatively from those for the
conventional ferromagnetic resonance   near thermal equilibrium and are amenable 
to experimental tests.

%In particular, we analyze fixed points
%of the spin vector in the plane transverse to the anisotropy axis, and study
%their stability under homogeneous and spin wave perturbations. We further
%obtain a generalized Suhl's instability criterion, giving the
%condition for exponential growth of P-modes under spin wave perturbations.

\end{abstract}

\end{frontmatter}

\section{Introduction}
Interest in ferromagnetic resonance (FMR) has soared in recent
times due to advances in fabricating nanostructures. This implies 
prospects for several new experiments to study possible absorption
phenomena, and patterns that may form owing to instabilities, in ferromagnetic
films. While other magnetic resonance counterparts such as nuclear magnetic
resonance(NMR), electron paramagnetic resonance(EPR), electron spin resonance, etc., 
have found immense technological applications, including crystal structure 
determination and medical diagnostics, FMR has remained a more complex 
phenomenon. Some of the main features of FMR are $a)$ large magnetization, 
and hence large susceptibility, $b)$ large demagnetization field due to strong
magnetization, which is also influenced by the shape of the specimen, and $c)$ 
resonance excitations breaking into spin wave modes, that make spin reversal 
more complex. 

Spin-wave instabilities were first observed independently by Damon \cite{damon}
and by Bloembergen and Wang \cite{bloem} as noisy anomalous absorption which 
abruptly sets in at a certain  threshold power as the resonant microwave field 
is increased. Suhl remarked that this phenomenon ``bears a certain resemblance 
to the turbulent state in fluid mechanics'' \cite{suhl} . In fact the 
instabilities in this case were first explored in \cite{suhl} and are referred 
to as {\it Suhl instability}. These were subsequently verified experimentally
\cite{hiller}.

Under a growing attention to deterministic chaos and nonlinear 
dynamics, there occurred a renaissance on studies involving high-power magnetic
resonance in the 
1980s. High-resolution experiments were carried out for spin-wave nonlinear 
dynamics in the yttrium iron garnet film and sphere in parallel and 
perpendicular pumpings \cite{resen,bryan1,bryan2,carrol}. For certain high 
powers beyond the Suhl threshold, interaction between excited spin-wave modes 
lead to various dynamical phenomena including auto-oscillations, 
period-doubling cascades, quasi-periodicity, and chaos. Also observed were 
irregular relaxation oscillations and abrupt transitions to wide-band 
turbulence, beyond the Suhl threshold \cite{resen,bryan1,bryan2,carrol}, 
which were explained by using Zakharov et al's {\it S-theory} \cite{lvov}. 
Since then the studies on spin wave instabilities have acquired  renewed
interests up to now \cite{wigen,beck,laul}.

Despite these pioneering works, including that of Suhl\cite{suhl}, the 
investigations have 
been limited to the instability around fixed points that correspond to  
magnetization parallel to the anisotropy axis. 
This is due to the following fact: In contrast to the case of NMR and EPR,
macroscopic ferromagnets in FMR have large frozen magnetization. Under a
static magnetic field, therefore, any available pumping field  cannot
freely rotate such strong magnetization at the cost of the large
stabilizing energy (Zeeman energy). In nanoscale ferromagnets, 
however,  it is possible to rotate the saturation magnetization far from the 
anisotropy axis. The magnetization can even be driven to the equatorial plane 
perpendicular to the anisotropy axis. Therefore it is timely to analyze the  
nonlinear dynamics of spin waves in these new non-equilibrium states.

It may also be noted that a number of investigations exist in the 
literature on the dynamics of higher dimensional Heisenberg ferromagnetic spin 
systems \cite{ml,kose,mike} corresponding to isotropic (pure exchange 
interaction), anisotropic, external field and other interactions 
\cite{ml,kose,mike,pod,arb}.  However, to the knowledge of the authors,
concerning   the systems coupled with Maxwell equation for the demagnetization
field as considered as in this paper, there exist very few studies.

In this article, we investigate a 2-D ferromagnetic system 
with uniaxial anisotropy in a background of alternating magnetic field 
transverse to, and a static magnetic field parallel to, the anisotropy axis. 
The axis of anisotropy is chosen to be arbitrary. We also 
include the demagnetization field due to the spatial inhomogeneity of the 
magnetization vector, which is seen to play a crucial role. Fixed points - 
P-modes - of such a system have been identified earlier, and their stability 
against both homogeneous perturbations and spin-wave excitations analyzed
\cite{bert1,bert2}. However, we specifically analyze equatorial   and other 
fixed points, that have not been identified earlier, wherein the average 
magnetization vector lies in the plane transverse to the anisotropy axis and 
exhibit a more complex dynamics. We further obtain a criterion for instability 
under spin wave excitations, thus generalizing the Suhl's instability criterion. 
 
  The strategy we follow in our analysis is as follows.  We project the unit 
spin vector $\spin{m}({\bf r},t)$ stereographically onto a complex plane 
$\psi({\bf r},t)$ and deduce the equation of motion in terms of the stereographic 
variable.  Now going over to a natural rotating frame of reference, one is able to 
identify the defining equations for the fixed points or the so called P-modes.  Of 
all the possible equilibrium points, the equatorial and related fixed points are of
special interest as their expressions can be explicitly obtained.  In fact we find 
that there exists four such equilibrium points.  Then we investigate their linear 
stability nature (i) under spatially homogeneous perturbations and (ii) under more 
general spin wave perturbation in order to identify their local and global 
stability, and obtain conditions for Suhl's instability as a function of 
experimentally measurable parameters such as the amplitudes of oscillatory 
perpendicular and static parallel external magnetic fields.  The results give clear
criteria for experimental realization of the predicted results.

The plan of the paper is as follows. In Section \ref{model}, we introduce the
model 
spin Hamiltonian for the ferromagnetic film in the presence of the external
field and an intrinsic demagnetization field, and write down the 
Landau-Lifshitz (LL) equation for the spin field with the Gilbert damping
term included. We further introduce the stereographic variable, and rewrite
the LL equation in terms of the new variable.
In Section \ref{fp}, we identify fixed points of the LL equation, specifically the 
equatorial and other  related fixed points, and analyze their linear 
stability under homogeneous perturbations. In Section \ref{efp}, we study the  
stability   of these fixed points under spin wave excitations in terms of a 
period map and generalize Suhl's instability criterion. 
We conclude with a summary of results in Section \ref{conc}. 
\section{\label{model}Model Hamiltonian and LL equation}
We consider a 2-D ferromagnetic film with uniaxial anisotropy and
an applied oscillating magnetic field in the direction transverse to the
anisotropy axis. Such a system can be described by a Hamiltonian for the unit
spin field 
$\spin{m}({\bf r},t)=\{m_1({\bf r},t),m_2({\bf r},t),m_3({\bf r},t)\};~~
\spin{m}^2=1$ as 

\beq\label{ham}
H = H_{exchange} + H_{applied} + H_{anisotropy} + H_{demagnetization},
\eeq
where
\beq\label{ham_e}
H_{exchange} = \int D(\nabla\spin{m})^2~d^2{\bf r},\\
\eeq
\beq
H_{applied} = -\int ~{\bf B}_a\cdot \spin{m}~d^2{\bf r}, 
\eeq
\beq
H_{anisotropy} = -\int~\kappa m_{\parallel}^2~d^2{\bf r},
\eeq
\beq\label{ham_eb}
H_{demagnetization} = -\int~{\bf H}_m\cdot\spin{m}~d^2{\bf r},
\eeq
and the demagnetizing field ${\bf H}_m$ satisfies the Maxwell's equation,
\beq\label{aux}
\nabla\cdot{\bf H}_m = -4\pi\nabla\cdot\spin{m};~~
\nabla\times{\bf H}_m = 0. 
\eeq
In the above, we have assumed the average magnetization to be of constant 
magnitude in space and time, and hence conveniently expressed the Hamiltonian 
(\ref{ham}) in terms
of the normalized unit average magnetization $\spin{m}$. Equation (\ref{aux})
gives the constitutive relation connecting the demagnetization field
with the average local magnetization, assuming absence of any free current in
the film. In equations (\ref{ham_e}-\ref{ham_eb}), D is the exchange constant, 
${\bf B}_a$
is the applied magnetic field, $\kappa$ is the anisotropy parameter and 
$m_{\parallel}$ is the component of magnetization in the direction of 
anisotropy.  

Time evolution of the spin field is governed by the Hamilton's 
equation 
\beq
\dot{\spin{m}}= \frac{\delta H}{\delta\spin{m}}=\{\spin{m},H\}.
\eeq 
Using the spin Poisson bracket relations 
\beq\label{comm}
\{m_i({\bf r},t),m_j({\bf r'},t')\}=\epsilon_{ijk}m_k\delta({\bf r-r'},t-t'),
\eeq 
it is straight forward to show that the dynamics is governed by the 
LL equation\cite{ll}, 
\beq\label{lle2}
\dot{\spin{m}}
= -\spin{m}\times{\bf F}_{eff},
\eeq
where ${\bf F}_{eff}$ is the effective field given by
\beq
\label{hef}
{\bf F}_{eff} = D\nabla^2\spin{m} + {\bf B}_a +\kappa m_{\parallel}\hat{e}_{\parallel} + {\bf H}_m. 
\eeq
In equation (\ref{hef}), $\nabla^2$ is the two dimensional Laplacian 
$\nabla^2=\partial^2/\partial x^2 +\partial^2/\partial y^2$. 
The direction of anisotropy $\hat{e}_{\parallel}$ has been chosen to be 
arbitrary in the internal space of the spin field. Further, writing 
$\hat{e}_{\parallel} = \{0,0,1\}$, we choose the applied 
magnetic field to be of the form
\beq\label{b_a}
{\bf B}_a = h_{a\perp}\{\cos\omega t, \sin\omega t, 0\} 
+h_{a\parallel}\hat{e}_{\parallel} 
\equiv h_{a\perp}(\cos\omega t\hat{e}_1 + \sin\omega t\hat{e}_2) 
+ h_{a\parallel}\hat{e}_{\parallel}, 
\eeq
where $\hat{e}_1$ and $\hat{e}_2$ are two unit orthonormal vectors in the
plane transverse to the anisotropy axis. Notice that the orientation of the 
anisotropy axis is specified as in equation (\ref{b_a}) in the spin space, but 
is still arbitrary in the physical space, 
where the plane of the ferromagnetic film is assumed to be in the $x-y$ 
direction. Without loss of generality one can choose $m_\parallel=m_3$. 
Introducing a phenomenological Gilbert damping term (proportional to $\alpha$),
equation (\ref{lle2}) can be modified to the form 
\beq\label{lle}
\dot{\spin{m}} - \alpha\spin{m}\times\dot{\spin{m}} 
= -\spin{m}\times{\bf F}_{eff}.
\eeq
It is this equation in 2-D   coupled with the Maxwell equation which we shall 
investigate in this paper.

It proves convenient for later analysis to express the equation in terms of the
complex stereographic variable   (see Figure 1 for details)
\beq\label{psi}
\psi \equiv \frac{m_1 + im_2}{1+m_3},
\eeq
or equivalently by the transformation
\beq
\spin{m} = \frac{1}{1+|\psi|^2}\{2Re\psi,2Im\psi,1-|\psi|^2\}.
\eeq
Correspondingly, the terms in the Hamiltonian in equation (\ref{ham}) become
\beq\label{ham_e2}
H_{exchange} = \int D\frac{|\nabla\psi|^2}{(1+|\psi|^2)^2}~d^2{\bf r},
\eeq
\beq
H_{applied} = -\int \Big[~h_{a\perp}\frac{\psi e^{-i\omega t}+\psi^*e^{i\omega t}}{1
+|\psi|^2} 
+ h_{a\parallel}\frac{(1-|\psi|^2)}{(1+|\psi|^2)}\Big]~d^2{\bf r},
\eeq
\beq
H_{anisotropy} = -\int~\kappa\frac{(1-|\psi|^2)^2}{(1+|\psi|^2)^2}~d^2{\bf r},
\eeq
\beq
H_{demagnetization} = -\int~\Big[\frac{H_m^*\psi+H_m\psi^*}{1+|\psi|^2}
+h_{m\parallel}\frac{(1-|\psi|^2)}{(1+|\psi|^2)}\Big]~d^2{\bf r},
\eeq
where we have used equation (\ref{b_a}), and represented the demagnetization field as
${\bf H}_m =\{h_1,h_2,h_{m\parallel}\}$ with $H_m=h_1+ih_2$.

On account of the spin Poisson bracket relations in equation (\ref{comm}), $\psi$ obeys
the Poisson bracket relation,
\beq\label{comm2}
\{\psi({\bf r},t),\psi^*({\bf r'},t')\}= i\delta({\bf r-r'},t-t').
\eeq
For the complex scalar field $\psi$, one can write down the evolution equation
from the Hamiltonian (\ref{ham}), equivalent to equation (\ref{lle2}). Then, including 
the damping term, the evolution equation for $\psi$ equivalent to equation (\ref{lle}) 
reads as \cite{naka} 
\bea\label{ll2}
i(1-i\alpha)\dot{\psi} = D(\nabla^2\psi
-2\frac{{\bPsi}(\nabla\psi)^2}{1+|\psi|^2})
-(h_{a\parallel} + \kappa\frac{1-|\psi|^2}{1+|\psi|^2})\psi\\\nonumber
\hskip 2.5cm +\frac{1}{2}h_{a\perp}(e^{i\omega t} - \psi^2e^{-i\omega t})\\\nonumber
\hskip 2.5cm -h_{m\parallel}\psi 
+ \frac{1}{2}(H_m -\psi^2H_m^*).
\eea

The form of the applied transverse field in equation (\ref{ll2}) suggests that it is
convenient to make a transformation 
\beq
\psi =\hsi({\bf r},t)\exp[{i\omega t}],
\eeq
and rewrite equation (\ref{ll2}) in terms of $\hsi$. This amounts to moving to a rotating
coordinate frame in the spin space, rotating about the anisotropy axis 
$\hat{e}_{\parallel}$ with angular frequency $\omega$. If $\spin{m}$ were 
expressed in spherical polar coordinates in the rotating frame as 
$\spin{m}=\{\sin\theta\cos\phi,\sin\theta\sin\phi,\cos\theta\}$, then 
$\hsi=\tan(\theta/2)\exp{i\phi}$. With this change, the equation (\ref{ll2}) becomes
\bea\label{llz3}
i(1-i\alpha)\dot{\hsi} = D(\nabla^2\hsi
-2\frac{\bhsi(\nabla\hsi)^2}{1+|\hsi|^2})
-(h_{a\parallel} -\omega +i\alpha\omega 
+ \kappa\frac{1-|\hsi|^2}{1+|\hsi|^2})\hsi\\\nonumber
\hskip 2.5cm +\frac{1}{2}h_{a\perp}(1 - \hsi^2)\\\nonumber
\hskip 2.5cm -h_{m\parallel}\hsi + \frac{1}{2}(H_me^{-i\omega t}
- \hsi^2H_m^* e^{i\omega t}). 
\eea
It must be remembered that $h_{m\parallel}, H_m$ and ${H}^*_m$ in equation (\ref{llz3})
are intrinsically related to $\hsi$ through the relation (\ref{aux}). 
Due to the second equation in (\ref{aux}), we can write
\beq
{\bf H}_m = -\nabla\varphi,
\eeq
by introducing an auxiliary potential $\varphi({\bf r},t)$. 
Thus, ${\bf H}_m$ can be found by solving the Poisson equation 
$\nabla^2\varphi = -4\pi\nabla\cdot\spin{m}$. Consequently, the boundary 
condition and shape 
of the boundary influence the actual form of $\varphi$, and hence ${\bf H}_m$. 
In the following we 
shall consider only spatially homogeneous spin fields and small amplitude spin 
waves. In this case, boundary effects can be ignored if one assumes the spin 
wave to be confined to a localized region, and that they dissipate close to
the boundary. In the case of large amplitude perturbations, coupling between
different modes become nontrivial. This case will be separately treated 
elsewhere. 
\section{\label{fp}Fixed points (P-modes) and stability}
Spatially homogeneous fixed points of equation (\ref{llz3}) correspond to a uniform
magnetization field that exhibit a periodic motion of frequency $\omega$, 
irrespective of the fact that the governing equations are highly nonlinear. 
Such homogeneous steady states in the rotating frame are referred to as 
P-modes. These fixed points $\hso$, in the moving frame, are obtained from 
equation (\ref{llz3}) by assuming time independence and spatial homogeneity of $\hsi$, 
i.e., $\partial{\hso}/\partial t=0=\nabla\hso$. 
For such a uniform steady field $\hso$, the uniform demagnetization field can 
be conveniently written as
${\bf H}^o_m = -N_{\perp}{\bf m}^o_{\perp}-N_{\parallel}{\bf m}^o_{\parallel}$,
where ${\bf m}^o_{\perp} = \{m^o_1,m^o_2,0\}$, i.e.,
\beq\label{H_m}
H^o_m = -N_\perp(m^o_1+im^o_2)=-2\frac{N_{\perp}\hso}{1+|\hso|^2}e^{i\omega t};
\eeq
\beq
h_{m\parallel} = -N_\parallel m^o_\parallel =
-N_{\parallel}\frac{1-|\hso|^2}{1+|\hso|^2}.
\eeq
Using these conditions in equation (\ref{llz3}), one obtains after some rearrangement,
the equation for the fixed points as
\beq\label{fixed}
(b_{a\parallel}+ i\Omega 
+ \frac{1-|\hso|^2}{1+|\hso|^2})\hso
-\frac{1}{2}b_{a\perp}(1 - \hso^2)=0,
\eeq
where the quantities $b_{a\parallel}, b_{a\perp}, \Omega$ and $\kappa_{eff}$ are
defined as 
\beq
b_{a\parallel} =  \frac{h_{a\parallel} -\omega}{\kappa_{eff}};
~~b_{a\perp} = \frac{h_{a\perp}}{\kappa_{eff}};
~~\Omega = \frac{\alpha\omega}{\kappa_{eff}};
~~\kappa_{eff} = \kappa+N_{\perp}-N_{\parallel}.
\eeq
Evidently, the fixed points are functions of the applied field strengths $h_{a\perp}$ 
and $h_{\parallel}$, and the frequency $\omega$, besides the anisotropy 
constant $\kappa$. However, of these $h_\parallel, h_{a\perp}$ and $\omega$ are the 
tunable parameters. 

From equation (\ref{fixed}), it is straightforward to check that the equation satisfied 
by $|\hso|^2$ is,
\beq\label{fixed2}
{|\hso|^2}\Big(b_{a\parallel} + \frac{1-|\hso|^2}{1+|\hso|^2}\Big)^2
+\Omega^2{|\hso|^2}\Big(\frac{1-|\hso|^2}{1+|\hso|^2}\Big)^2
=\frac{b_{a\perp}^2}{4}{(1 - |\hso|^2)^2}.
\eeq
Equation (\ref{fixed2}) can be rewritten as a polynomial of degree four in $u=|\hso|^2$
as
\bea\label{quartic}
-\frac{b_{a\perp}^2}{4}u^4+(b_{a\parallel}^2-2b_{a\parallel}+1+\Omega^2)u^3
+(2b_{a\parallel}^2-2-2\Omega^2+\frac{b_{a\perp}^2}{2})u^2\\\nonumber
\hskip 2cm +(b_{a\parallel}^2+\Omega^2+1+2b_{a\parallel})u
-\frac{b_{a\perp}^2}{4}=0.
\eea
The quartic equation (\ref{quartic}) admits in general zero, two or four real
solutions depending on the sign and magnitude of the coefficients. These roots can 
be easily obtained for any given set of parameters, and the nature of the 
corresponding fixed points can
be analyzed by means of standard theory of equilibrium points of 
two-dimensional dynamical systems. One can establish the presence of either
stable modes or foci, unstable nodes or foci, or saddles. A detailed analysis
of these modes in terms of polar coordinates has been made in the past by 
Bertotti et. al. \cite{bert1}.

In this paper, we wish to concentrate on certain physically interesting modes
not identified earlier,
namely those arising out of fixed points along the equator   or related fixed
points, whose form can
be explicitly determined. Consequently their stability analysis can also be 
carried out in detail. In particular,
on account of the anisotropy, the tendency for the spin field is to orient
along the direction of $\pm\hat{e}_\parallel$. Naturally, the fixed points 
along the equator,
i.e., in the $\hat{e}_1-\hat{e}_2$ plane, are special, being the other 
extreme,  but are available in nonequilibrium nano-ferromagnets. Besides, as will be shown below, these fixed points are degenerate, 
and the nature of dynamics around these fixed points is naturally of 
considerable 
interest. In the next three sections, we identify these fixed points and study
their stability under homogeneous and spin wave perturbations. In Section 
\ref{suhl},
we obtain the generalization of the Suhl's instability criterion. 

\subsection{Equatorial and related fixed points}
\label{hom_f}
In this section we identify fixed points along the equator   and others related
to them, and analyze
their stability under homogeneous perturbations. 

Along the equator, we have $m^o_\parallel(=m^o_3)=0$, or $|\hso|^2=1$. 
Conversely, $|\hso|^2=1$ implies spins are pointed in the equatorial plane. 
But from equation (\ref{fixed2}),fixed points along the equator are possible only 
when $b_{a\parallel}=0$,or $h_{a\parallel}=\omega$. However for this choice of the 
pumping frequency $\omega$ or the  equivalent amplitude of the static parallel 
field 
$h_{a\parallel}$, other related fixed points also exist.  In fact with this choice,
equation (\ref{fixed}) can be solved to obtain four fixed points $P_1, P_2,P_3$ and
$P_4$:
\beq\label{phi_o}
\frac{\hsi_{oIm}}{\hsi_{oRe}} = \tan\phi = -\Omega;
\eeq
\beq\label{mod}
P_1=P_2 : |\hso|^2=1;
\eeq
\beq
P_3: |\hso|^2 = \frac{2(1+\Omega^2)}{b_{a\perp}^2}\bigg[1+\sqrt{1
    -\frac{b_{a\perp}^2}{1+\Omega^2}}\bigg] -1;
\eeq
\beq\label{modb}
P_4: |\hso|^2 = \frac{2(1+\Omega^2)}{b_{a\perp}^2}\bigg[1-\sqrt{1
    -\frac{b_{a\perp}^2}{1+\Omega^2}}\bigg] -1.
\eeq
The fixed points are adjusted by tuning the transverse field strength 
$h_{a\perp} (=\kappa_{eff}b_{a\perp})$ and the frequency 
$\omega(=\kappa_{eff}\Omega/\alpha)$
of the applied field, while keeping $h_{a\parallel}=\omega$. Equations 
(\ref{mod})-(\ref{modb}) also impose a limit on the 
applied transverse field strength, namely, $0\le b_{a\perp}^2\le 1+\Omega^2$. 
  In terms of the original variables this condition becomes $0\le h_{a\perp}^2 
\le \kappa_{eff}^2+\alpha^2\omega^2$.

From equation (\ref{fixed2}) we notice that there are four solutions to 
$|\hso|^2$ as seen above. Of these
four, two fixed points, $(P_1,P_2)$ merge $(|\hso|^2=1)$ with the choice 
$b_{a\parallel}=0$, or $h_{a\parallel}=\omega$. These fixed points lie along the 
equator, transverse to 
$\hat{e}_\parallel$. The other two   related fixed points $(P_3,P_4)$ in 
equations (\ref{mod}-\ref{modb}) vary 
between south pole $(|\hso|^2=\infty)$ and equator for the range 
$0\le b_{a\perp}^2\le 1+\Omega^2$. The exact orientation will also be decided 
by the azimuthal angle $\phi$ determined by $\Omega$, as in equation (\ref{phi_o}). 
Particularly, for the choice $b_{a\perp}^2= 1+\Omega^2$,  i.e., 
$h_{a\perp}^2=\kappa_{eff}^2+\alpha^2\omega^2$,  the only possible
fixed points are along the equator. 
\subsection{Spatially homogeneous perturbations}
For a spatially homogeneous perturbation $\dsi(t)$ from $\hso$, the linear 
stability of $\hso$ 
is analyzed using equation (\ref{llz3}), with the assumption 
$\hsi =\hso+\epsilon\dsi, \epsilon <<1$, and restricting to linear
terms in $\dsi$. Correspondingly, the perturbed equation reads as 
\bea\label{h_per}
i(1-i\alpha)\delta\dot{\psi} = -\kappa_{eff}( i\Omega 
+ \frac{1-|\hso|^2}{1+|\hso|^2})\dsi
+2\kappa_{eff}\Gamma\hso(\hso\dsi^*+\bhso\dsi)\\\nonumber
-h_{a\perp}\hso\dsi, 
\eea
where we have defined $\Gamma = (1+|\hso|^2)^{-2}$. 
The equation (\ref{h_per}) and its complex conjugate can be written in a compact
form using an appropriate matrix ${\bf \Lambda}$, as 
\beq
\frac{d\Psi}{dt} = \Lambda\cdot\Psi,
\eeq
with 
\beq
\mathbf{\Lambda}= 
\left(\begin{array}{cc}
\frac{i\kappa_{eff}}{1-i\alpha}
\Big[i\Omega + \frac{1-|\hso|^2}{1+|\hso|^2}-2\Gamma|\hso|^2 &
\frac{-2i\kappa_{eff}}{1-i\alpha}\Gamma\hso^2\\ 
\hskip 1cm +b_{a\perp}\hso\Big] & \\
&\frac{-i\kappa_{eff}}{1+i\alpha}
\Big[-i\Omega + \frac{1-|\hso|^2}{1+|\hso|^2}-2\Gamma|\hso|^2\\
\frac{2i\kappa_{eff}}{1+i\alpha}\Gamma{\hso}^{*2} 
&\hskip 1cm +b_{a\perp}\hso^*\Big]
\end{array}\right)
\eeq
and $\Psi = (\dsi,\dsi^*)^T$. The type of instability of a 
fixed point is determined by the determinant and trace of ${\bf\Lambda}$. 
Defining $\chi=(1-|\hso|^2)/(1+|\hso|^2)$, from 
equation (\ref{h_per}), the determinant and trace are found to be
\beq
|{\bf \Lambda}| = \frac{\kappa_{eff}}{1+\alpha^2}\chi^2(1+\Omega^2),
\eeq
\beq
Tr({\bf \Lambda}) = -2\alpha\frac{\kappa_{eff}}{1+\alpha^2}(
\frac{1+\chi^2}{2} +\frac{\chi\Omega}{\alpha}).
\eeq
In general, the fixed point is a stable node or foci if $|{\bf \Lambda}|>0$ and
$Tr{\bf \Lambda}<0$, an unstable node or foci in case $|{\bf \Lambda}|>0$ and 
$Tr{\bf \Lambda}>0$, and a saddle if $|{\bf \Lambda}|<0$. However, for the 
equatorial fixed points $|\hso|^2=1$, and consequently $|{\bf \Lambda}|=0$,
since $\chi=0$, thus leading to degeneracy. 
   In order to identify the nature of these fixed points, we directly
solve equation (\ref{llz3}) numerically in the neighborhood of the fixed points,
assuming spatial inhomogeneity of $\hsi$ and plot the resultant phase portrait
in Figure 2. 
%The phase portraits in Figure 2, obtained using equation (\ref{llz3}), reveals 
%the nature of dynamics around these fixed points. 
Figure 2(a) shows the
phase portrait for the choice $b_{a\perp}^2=1+\Omega^2$, for which the only 
fixed points are along the equator $(P_1 \rm{and} P_2)$, as remarked earlier. 
Figure 2(b) is for the choice $b_{a\perp}^2=(1+\Omega^2)/2$, for which
there are two   additional fixed points at $\hso=(2.4,-0.2)(P_3)$ and 
$\hso=(0.41,-0.03)(P_4)$.  
\section{\label{efp}Equatorial and related fixed points and spin-wave instability}
Unlike in EPR and NMR, 
spin reversal is prevented in ferromagnetic resonance because of   the large
frozen magnetization and of the emergence
spin wave modes coupling to resonance excitations. 
In this section we generalize the study in \cite{suhl} to the case of  equatorial
states of ferromagnet with uniaxial anisotropy in an arbitrary direction, and 
a demagnetization field. In Section \ref{4.2} we concentrate specifically on 
equatorial fixed points in the presence of spin wave excitations. In Section 
\ref{suhl} we obtain the generalized Suhl's instability criterion. 
For this purpose, we consider first a general perturbation $\dsi({\bf r},t)$ to
the homogeneous steady field $\hso$, in contrast to the spatially homogeneous
perturbations of the previous section. The nature of instabilities are 
identified by linearizing equation (\ref{llz3}) in $\dsi$, around $\hso$.
Upon substituting $\hsi\equiv\hso+\dsi$ in equation (\ref{llz3}), and confining to
first order in $\dsi$, we find,
\bea\label{LLZ}
i(1-i\alpha)\delta\dot{\hat{\psi}} = D\nabla^2\dsi
-(h_{a\parallel} -\omega +i\alpha\omega 
+ \kappa\frac{1-|\hso|^2}{1+|\hso|^2})\dsi\\\nonumber
\hskip 2.5cm +2\kappa\Gamma\hso(\hso\dsi^*+\bhso\dsi)\\\nonumber
\hskip 2.5cm -h_{a\perp}\hso\dsi -\delta h_{m\parallel}\hso\\\nonumber
\hskip 2.5cm + \frac{1}{2}\delta H_me^{-i\omega t} 
-\frac{1}{2}\hsi^2\delta{H}^*_me^{i\omega t}. 
\eea
Note that in the above equation (\ref{LLZ}), the dispersion term and the variations in 
the demagnetization field are nontrivial due to the perturbation being 
spatially inhomogeneous. Making now a spatial Fourier decomposition,
\beq\label{dsi}
\dsi = \sum_{\bf q} {a}_q(t) e^{i{\bf q\cdot r}};~~~
\delta\bPsi = \sum_{\bf q} {a}^*_{-q}(t) e^{i{\bf q\cdot r}},
\eeq
and substituting in equation (\ref{LLZ}) we obtain the mode equations for ${a}_q$
 with the propagation vector ${\bf q}$. 
Due to the Poisson bracket relations in equation (\ref{comm2}), the mode amplitudes 
${a}_q$ also obey the Poisson bracket relation 
$\{{a}_q,{a}^*_{-q'}\}=i\delta_{q,q'}$. 

The corresponding change in the demagnetization field is obtained using
the relation 
\beq
\nabla\cdot\delta{\bf H}_m = -4\pi\nabla\cdot\delta{\bf m}.
\eeq
Here, $\delta{\bf m}$ is the change in the average local magnetization
corresponding to $\dsi$. This can be calculated using equation (\ref{psi}), and is given 
by
\bea\label{dm}
\delta{\bf m} = \Gamma\bigg\{
(e^{-i\omega t}-\hso^2e^{i\omega t})\bdsi +
(e^{i\omega t}-\hso^{*2}e^{-i\omega t})\dsi,\\\nonumber
\hskip 1.75cm i(e^{-i\omega t}+\hso^2e^{i\omega t})\bdsi -
i(e^{i\omega t}+\hso^{*2}e^{-i\omega t})\dsi,\\\nonumber
\hskip 1.75cm -2(\hso\bdsi + {\bhso}\dsi)
\bigg\}.
\eea
Using equation (\ref{dsi}) in equation (\ref{dm}) we can write,
\beq
\delta{\bf m}  = \sum_{\bf q} {\bf Z}_q(t) e^{i{\bf q\cdot r}},
\eeq
where
\bea
{\bf Z}_q(t) = \Gamma\bigg\{
{{a}^*}_{-q}(e^{-i\omega t}-\hso^2e^{i\omega t}) +
{{a}}_{q}(e^{i\omega t}-{\bhsos}e^{-i\omega t}), \\\nonumber
\hskip 2.5cm i{{a}^*}_{-q}(e^{-i\omega t}+\hso^2e^{i\omega t}) -
i{{a}}_{q}(e^{i\omega t}+{\bhsos}e^{-i\omega t}), \\\nonumber
\hskip 2.5cm -2(\hso{{a}^*}_{-q}+\bhso{a}_q)\bigg\}
\eea
\beq
\hskip 1.8cm = {\bf Z}^*_{-q}(t).
\eeq
Consequently, it can be verified that
\beq\label{dHm}
\delta{\bf H}_m = -4\pi\sum_{\bf q}\frac{{\bf q}}{q^2}
({\bf q\cdot Z}_q)e^{i{\bf q\cdot r}}
\eeq
satisfies the constitutive relations in equation (\ref{aux}). 

From equation (\ref{LLZ}), using equations 
(\ref{dsi}) and (\ref{dHm}), we have the mode equation satisfied by the
amplitudes ${a}_q$,
\bea\label{mode}
i(1-i\alpha)\dot{a}_q = \bigg[
-Dq^2-\Big(h_{a\parallel} -\omega +i\alpha\omega 
+ \kappa\frac{1-|\hso|^2}{1+|\hso|^2}\Big)\bigg]{a}_q\\\nonumber
\hskip 2.5cm +2\kappa\Gamma\hso(\hso{a}^*_{-q}+\bhso{a}_q)\\\nonumber
\hskip 2.5cm -h_{a\perp}\hso{a}_q 
-\delta h_{m\parallel}\hso\\\nonumber
\hskip 2.5cm + \frac{1}{2}\delta H_me^{-i\omega t} 
-\frac{1}{2}\hso^2\delta{H}^*_me^{i\omega t},
\eea
where $\delta H_m$ and $\delta h_{m\parallel}$ are obtained from equation (\ref{dHm}). 
If ${\bf \hat{q}(\equiv q/|q|)}$ is expressed in the spin frame as 
${\bf\hat{q}}=\sin\theta_q(\cos\phi_q\hat{e}_1+\sin\phi_q\hat{e}_2) 
+\cos\theta_q\hat{e}_{\parallel}$, then
\bea\label{qAq}
{\bf\hat{q}\cdot Z}_q = \Gamma\bigg[
\sin\theta_q\Big[{a}_q(e^{i(\omega t-\phi_q)}
-\bhsos e^{-i(\omega t-\phi_q)})\\\nonumber
\hskip 3.5cm +{a}^*_{-q}(e^{-i(\omega t-\phi_q)}
-\hso^2 e^{i(\omega t-\phi_q)})\Big]\\\nonumber
\hskip 2cm -2\cos\theta_q(\hso{a}^*_{-q}+\bhso{a}_q)\bigg]
\eea
\beq
={\bf\hat{q}\cdot Z^*}_{-q}.
\eeq
  Here $\theta_q$ and $\phi_q$ are the polar and azimuthal angles made
by the wave vector ${\bf q}$ in the spin frame spanned by $\hat{e}_1$, $\hat{e}_2$
and $\hat{e}_{\parallel}$. 
Substituting equation (\ref{qAq}) in equation (\ref{dHm}) and using the 
resultant expression for $\delta{\bf H}_m$ in equation (\ref{mode}), we can 
write 
\beq\label{mode1}
\dot{a}_q +i{A}_q {a}_q +i{B}_q {a}^*_{-q}=0,\\
\eeq
\bea\label{a_q}
{A}_q = (1-i\alpha)^{-1}\bigg[-Dq^2-\Big(h_{\parallel}-\omega +i\alpha\omega+
\kappa\frac{1-|\hso|^2}{1+|\hso|^2}\Big)\\\nonumber
\hskip 3cm + 2\kappa\Gamma|\hso|^2-h_{a\perp}\hso\\\nonumber
\hskip 3cm -2\pi\Gamma|2\hso\cos\theta_q-\sin\theta_q(e^{-i(\omega t-\phi_q)}
-\hso^2 e^{i(\omega t-\phi_q)})|^2\bigg]\\\nonumber
\eea
\bea\label{b_q}
{B}_q = (1-i\alpha)^{-1}\bigg[2\kappa\Gamma\hso^2 
-2\pi\Gamma\Big(2\hso\cos\theta_q\\\nonumber
\hskip 3cm -\sin\theta_q(e^{-i(\omega t-\phi_q)}
-\hso^2 e^{i(\omega t-\phi_q)})\Big)^2\bigg].
\eea
Similarly,
\beq\label{mode12}
\dot{a}^*_{-q} -i{A}^*_{-q} {a}^*_{-q} +i{B}^*_{-q} {a}_{q}=0.\\
\eeq

Now we will analyze these mode equations (\ref{mode1}) and (\ref{mode12}) in some 
detail in the following subsections.
\subsection{\label{4.2}Stability of equatorial   and related} fixed points
As we noted in Section \ref{hom_f}, the fixed points along the equator,
  $P_1$ and $P_2$, as well as the related fixed points $P_3$ and $P_4$ are 
possible when the pumping frequency $\omega$ coincides with the amplitude of the
static parallel magnetic field $h_{a\parallel}$.   The nature of stability of 
these fixed points can be studied by identifying a period map ${\bf M}$ using the 
mode equations (\ref{mode1}) and (\ref{mode12}). In this connection it must be 
remembered 
that the P-modes have inherent in them a time period $T=2\pi/\omega$ arising 
from the frequency of the applied transverse field.   Then the period map 
${\bf M}$, defined through the equation,
\beq\label{M}
\left(\begin{array}{c}
{a}_q(T)\\
{a}^*_{-q}(T)
\end{array}\right)
=\mathbf{M}
\left(\begin{array}{c}
{a}_q(0)\\
{a}^*_{-q}(0)
\end{array}\right),
\eeq
is obtained from equations (\ref{mode1}) and (\ref{mode12}) after integration through
one period $T$. Numerically, the map ${\bf M}$ in 
equation (\ref{M}) is obtained by evolving over a period $T$ the two column vectors 
$(1,0)^T$ and $(0,1)^T$, taken as the initial vectors $(a_q(0),a^*_{-q}(0))^T$.
The two columns   so obtained then form the columns of the period map ${\bf M}$.
From equations (\ref{phi_o}) and (\ref{mod})-(\ref{modb}), there are effectively two 
tunable parameters, $\Omega$ and $b_{\perp}$   or $\omega$ and $h_{a\perp}$, that
also fix the fixed point $\hso$. 
If $\gamma_{q\pm}$ are the eigenvalues of the period map ${\bf M}$ for a given
fixed point $\hso$, the fixed point is unstable for a mode of wave vector 
${\bf q}$ if $|\gamma_{q\pm}|>1$. Figure 3 shows unstable regions in the 
$\cos\theta-\omega$ space as shaded regions for different values of the 
angle $\theta_q$: (a) $\sin\theta_q=0$, (b) $\sin\theta_q=0.6$ and (c)
$\sin\theta_q=1$.   Note that $\cos\theta$ here effectively stands for 
$h_{a\perp}$ through the relations (\ref{mod})-(\ref{modb}). 
  In Figure 4, the same regions are shown in the 
$h_{a\perp}-h_{a\parallel}$ plane. However, not all points on the plane are
experimentally accessible if one requires a homogeneous background, due to the
condition $0\le h_{a\perp}^2 \le \kappa_{eff}^2+\alpha^2\omega^2$.
Note that $\theta_q$ defines the angle between the anisotropy
axis and the plane of the ferromagnetic film. 
 It should be emphasized that the global stability diagrams in Figures 3 and 4
are obtained under the condition that $\omega=h_{a\parallel}$. We find that the 
equatorial and related fixed points are always  stable for the perpendicular field
with its strength $h_{a\perp}$ less than the value determined by the pumping frequency,
but become unstable when $h_{a\perp}$ is ``decreased'' below the generalized Suhl 
threshold. There exists an interesting singular structure near the vicinity of
$\omega=0$ (Figure 3) or of $h_{a\parallel}=0$ (Figure 4), which
is very sensitive to the angle $\theta_q$. These phase diagrams differ
qualitatively from those for the conventional spin-wave instability
near the ferromagnetic ground state.

\subsection{\label{suhl} Mode equations and Suhl's instability}
It can be verified from equations (\ref{a_q}) and (\ref{b_q}) that 
${A}_q={A}_{-q}$ and ${B}_q={B}_{-q}$, if we note 
${\bf\hat{q}\to -\hat{q}} \Rightarrow \theta_q\to\pi-\theta_q$ and 
$\phi_q\to\phi_q+\pi$.
To begin with, we shall first suppress the oscillating terms with 
$exp[\pm i(\omega t -\phi_q)]$ and $exp[\pm 2i(\omega t -\phi_q)]$. 
Then, ${A}_q$ and ${B}_q$ in equation (\ref{mode1}) are replaced by $\taq$ 
and $\tbq$ given by 
\bea\label{a2_q}
\taq = (1-i\alpha)^{-1}\bigg[-Dq^2-\Big(h_{\parallel}-\omega +i\alpha\omega+
\kappa\frac{1-|\hso|^2}{1+|\hso|^2}\Big)\\\nonumber
\hskip 3cm + 2\kappa\Gamma|\hso|^2-h_{a\perp}\hso\\\nonumber
\hskip 3cm -2\pi\Gamma\Big(
4|\hso|^2\cos^2\theta_q+\sin^2\theta_q(1+|\hso|^4)\Big)\bigg]
\eea
\bea\label{b2_q}
\tbq = (1-i\alpha)^{-1}\bigg[2\kappa\Gamma\hso^2 
-2\pi\Gamma\Big(4\hso^2\cos^2\theta_q
-2\sin^2\theta_q\hso^2\Big)\bigg]
\eea
Then equation (\ref{mode}) can be written in terms of normal modes, after a 
Bogoliubov type linear 
transformation ${a}_q = \lambda_q{b}_q -\mu_q{b}^*_{-q}$, to a
new variable ${b}_q$. Here,
on account of both ${a}_q$ and ${b}_q$
obeying Poisson bracket relations, $\lambda_q$ and $\mu_q$ satisfy
the condition,
\beq\label{norm}
\lambda_q\lambda_{-q}-\mu_q\mu^*_{-q}=1.
\eeq
However, $\lambda_q=\lambda_{-q}$ and $\mu_q=\mu_{-q}$, as will be shown below.
With this change, using equation (\ref{norm}), equation (\ref{mode1})  becomes
\bea\label{mode2}
{\dot{b}}_q + i\Big[\lambda_q(\lambda_q\taq-\mu^*_{q}\tbq) 
+\mu_q(\mu^*_{q}\taq^*-\lambda_q\tbq^*)\Big]{b}_q
+i\Big[\lambda_q(\lambda_q\tbq-\mu_q\taq)\\\nonumber
\hskip 3cm 
+\mu_q(\mu_q\tbq^*-\lambda_q\taq^*)\Big]{b}^*_{-q} =0.  
\eea
In order for ${b}_q$ to represent normal modes, the last 
term in equation (\ref{mode2}) must vanish. This condition along with equation (\ref{norm}) 
determines $\lambda_q$ and $\mu_q$. After some straight forward algebra we find
\beq\label{eta}
\lambda_q = \cosh\frac{\xi_q}{2};
~~~~\mu_q = \sinh\frac{\xi_q}{2}e^{i\nu_q},
\eeq
\beq\label{nu}
\tanh\xi_q = \frac{2|\tbq|}{\taq+\taq^*};
~~~~e^{2i\nu_q} = \frac{\tbq}{\tbq^*}.
\eeq
Since $\taq=\tilde{A}_{-q}$ and $\tbq=\tilde{B}_{-q}$ from equations 
(\ref{a2_q}) and (\ref{b2_q}), it follows that $\lambda_q=\lambda_{-q}$ and 
$\mu_q=\mu_{-q}$. Substituting for $\lambda_q$ and $\mu_q$ from equations 
(\ref{eta}) and (\ref{nu}) in equation (\ref{mode2}) gives,
\beq\label{normal}
\dot{b}_q +i(\omega_q - i\eta_q){b}_q =0,
\eeq
\beq
\omega^2_q = \frac{1}{2}[(\taq+\taq^*)^2-4|\tbq|^2];
\eeq
\beq
\eta_q = \frac{i}{2}(\taq-\taq^*),
\eeq
with solutions
\beq\label{bq_sol}
{b}_q \propto e^{-i\omega_qt-\eta_qt}.
\eeq
In equation (\ref{bq_sol}), $\omega_q$ is the eigen frequency for the spin-wave normal 
mode and $\eta_q$ is its damping constant. 

Next we analyze the spin wave dynamics in the presence of oscillating terms
in equation (\ref{mode1}). We notice from equations (\ref{a_q}) and (\ref{b_q}) that 
terms with $\exp[\pm i\omega t]$ and $\exp[\pm 2i\omega t]$ appear in 
equation (\ref{mode1}). First, we
suppress all other oscillating terms retaining only the term with 
$\exp[-2i\omega t]$. The equation (\ref{normal}) is modified 
to 
\beq\label{fre2}
 \dot{b}_q +i(\omega_q - i\eta_q){b}_q 
+ie^{-2i\omega t}\rho{b}^*_{-q}=0,
\eeq
where 
\beq\label{critpara}
\rho = \lambda_q(\lambda_q\tbq^{1}{''}-\mu_q\taq{''})
+\mu_q(\mu_q\tbq^{2}{''}-\lambda_q\taq{''}^*),
\eeq
\beq
\taq{''}\equiv \frac{2\pi\Gamma}{1-i\alpha}\sin^2\theta_q\hso^2.
\eeq
\beq
\tbq^1{''}\equiv \frac{-2\pi\Gamma}{1-i\alpha}\sin^2\theta_q\hso^4,
\eeq
\beq\label{critparb}
\tbq^2{''}\equiv \frac{-2\pi\Gamma}{1+i\alpha}\sin^2\theta_q.
\eeq
If we assume solutions of the form 
${b}_q = {b}^o_q(t)e^{-i\omega t -\eta_q}$, equation (\ref{fre2}) can be written 
as
\beq
\dot{b}^o_q +i(\omega_q - \omega){b}^o_q +\rho{b}^{o*}_{-q}=0,
\eeq
or, taking a second derivative in time, and using its complex conjugate,
\beq
\Big[\partial^2_t + (\omega_q-\omega)^2 - |\rho|^2\Big]{b}^o_q =0.
\eeq
The necessary condition to see a growth of spin waves against P-mode is
then given by
\beq
(\omega_q-\omega)^2 - |\rho|^2 \le 0.
\eeq
When this condition is satisfied, the normal mode grows as
\beq
{b}_q \propto \exp\Big[(|\rho|^2-(\omega_q-\omega)^2)^{1/2}t\Big]~
\exp[-i\omega_q t-\eta_q t].
\eeq
Thus, to see an exponential catastrophic growth of ${b}_q$, we should
have 
\beq\label{crit}
(\omega_q-\omega)^2 +\eta_q^2\le |\rho|^2. 
\eeq

With a similar argument for the term $e^{-i\omega t}$, repeating equations
(\ref{fre2}) - (\ref{crit}) we obtain a condition analogous to equation (\ref{crit}),
\beq\label{crit2}
(\omega_q-\frac{\omega}{2})^2 +\eta_q^2\le |\tilde{\rho}|^2, 
\eeq
with 
\beq\label{crit2para}
\tilde\rho = \lambda_q(\lambda_q\tbq^1{'}-\mu_q\taq{'})
+\mu_q(\mu_q\tbq^2{'}-\lambda_q\taq{'}^*),\\
\eeq
\beq
\taq{'} \equiv \frac{-2\pi\Gamma}{1-i\alpha}2\sin\theta_q\cos\theta_q(|\hso|^2-1)\hso,
\eeq
\beq
\tbq^1{'} \equiv \frac{-2\pi\Gamma}{1-i\alpha}4\sin\theta_q\cos\theta_q\hso^3,
\eeq
\beq\label{crit2parb}
\tbq^2{'} \equiv \frac{2\pi\Gamma}{1+i\alpha}4\sin\theta_q\cos\theta_q\hso^*.
\eeq
In the above analysis, the angle $\theta_q$ - the angle between the anisotropy 
axis and the $x-y$ plane plays a crucial role. From equations (\ref{a_q}) and
(\ref{b_q}), we note that when $\theta_q=\pi/2$, terms with 
$exp[\pm i\omega t]$ vanish. Consequently, the resonance at $\omega/2$ is 
absent. When $\theta_q =0$, the anisotropy axis is in the $x-y$ 
plane, and both the resonances $(\omega_q=\omega/2,\omega)$ vanish.  
Equations (\ref{crit}) with equations (\ref{critpara})-(\ref{critparb}), and 
equation (\ref{crit2}) with equations (\ref{crit2para})-(\ref{crit2parb})
provide new criteria for spin-wave instabilities of  P-modes.  The global
stability diagrams depicted on the basis of this new criteria are given in
Figure 5, which nicely characterize the principal part of the boundary curves
between stable and unstable regions in Figure 3 and 4.
 In particular, consistent with Figures 3 and 4 obtained from
the period map, there is a singular structure near the vicinity of
$\omega=0$ (Figure 5(b'), 5(c')) or
of $h_{a\parallel}=0$ (Figure 5(b") and 5(c")) sensitive to the angle
$\theta_q$.

\section{\label{conc}Conclusion}
We have investigated non-equilibrium states of nanoscale ferromagnets with
uniaxial anisotropy, in the presence of an oscillating field transverse to
the axis of anisotropy. The saturation magnetization can be driven even to
the equatorial plane perpendicular to the anisotropy axis. The P-modes
correspond to new non-equilibrium states lying far from the anisotropy
axis. The stability of the P-modes under uniform and spin wave
perturbations has been studied in general. 

%In particular, we have explored the stability of equatorial and related 
%non-equilibrium states of the nano-ferromagnet in the presence of perpendicular 
%pumping.  

  Specifically we have identified new equatorial and other fixed points, which 
exhibit a more complex dynamical phenomena. These novel states (fixed points) are 
realized in the case that the pumping frequency 
coincides with the amplitude of the static parallel field.  We concentrated on the 
spin-wave instability of the above states and found that these states are always 
stable for the perpendicular field with its strength $h_{a\perp}$ less than the 
value determined by the pumping frequency, but become unstable when $ h_{a\perp}$ is
``decreased'' below the generalized Suhl threshold.  By using the period map the 
global stability diagram in $ h_{a\perp}-h_{a\parallel}$ plane is obtained, which 
shows an interesting   singular structure which is very sensitive to the angle 
$\theta_q$ between 
the spin-wave propagation vector (the film plane) and the direction of the uniaxial 
anisotropy.

%From the linear stability analysis, we also notice that resonance occurs
%at the frequencies $\omega_q=\omega/2$ and $\omega_q=\omega$.  The resonances are 
%absent  when the axis of uniaxial anisotropy lies in the plane of the ferromagnet.
%when the wave vector associated with the spin wave lies in the x-y plane, 
%which is the plane of the ferromagnetic film. 

Closely following the original  paper 
by Suhl based on Bogoliubov-type transformation, we have also obtained the 
generalized Suhl instability condition in a form of coupled spin-wave equations,
 which recovers the major part of the global phase diagram obtained by the
period map.   It should be noticed that resonance occurs at the frequencies 
$\omega_q=\omega/2$ and $\omega_q=\omega$ for $\theta_q\neq 0$, but is absent for  
$\theta_q=0$, i.e., when the axis of uniaxial anisotropy lies in the plane of the 
ferromagnet.
The present findings differ qualitatively from those for the conventional
  Suhl instability near the ground state of uniaxial ferromagnets.
Various nonlinear dynamics aspects including spin wave turbulence (see for 
example ref. \cite{arb2,cross}) beyond the generalized Suhl instability
will constitute subjects which we intend to study in future.

%Further, we have obtained generalized Suhl's spin-wave
%instability criterion in the presence of dissipation, anisotropy in an
%arbitrary direction, and demagnetization. From the linear stability
%analysis, we notice resonance occurs at the frequencies
%$\omega_q=\omega/2,\omega$. The arbitrariness of the axis of anisotropy is
%crucial in determining the resonance frequencies. The resonances are
%absent when the wave vector associated with the spin wave lies in the
%$x-y$ plane, the plane of the ferromagnetic film. Various nonlinear
%dynamics including spin-wave turbulence beyond the generalized Suhl
%instability will constitute subjects which we intend to study in future.

\ack
CK and KN are grateful to JSPS for the financial support of the Fundamental
Research, C-2, No. 16540347, entitled `Unified Approach to Quantum Chaos and 
Macroscopic Quantum Dynamics'.
The work of SM and ML forms part of a project sponsored by the Department of 
Science and Technology, Government of India.

\newpage
\begin{figure}[h]
%\centering\includegraphics[width=.8\linewidth]{Figure1.eps}
\caption{Stereographic projection from the unit sphere onto a complex plane.
Any point ${\bf\hat{m}}$ on the surface of the sphere can be mapped onto a 
point $\psi$ on the complex plane as shown above. Points on the upper
hemisphere are mapped   onto points inside the unit circle, while points on the 
lower hemisphere are mapped to the points outside the unit circle. All 
points at $\infty$ in the complex plane are identified with the south pole
${\bf S}$.} 
\end{figure}

\begin{figure}[h]
%\centering\includegraphics[width=1.2\linewidth]{Figure2.eps}
\caption{Phase portrait obtained using equation $(22)$, assuming spatial 
homogeneity of $\hsi$, for the choice $h_{a\parallel}=\omega$, 
$\kappa_{eff}=0.1$, $D=1$, $\alpha=0.01$ and $\omega=0.8$. The initial value
of Im$\hat{\psi}$ is fixed at $0.5$, while Re$\hat{\psi}$ is 
varied from $-3$ to $5$. (a) $b^2_{a\perp}=1+\Omega^2$, with fixed points 
corresponding to $|\hso|^2=1$,$P_1=(-0.99,0.12)$ (unstable
fixed point), and $P_2=(0.99,-0.12)$ (stable fixed point). 
(b) $b^2_{a\perp}=(1+\Omega^2)/2$, with fixed points at $(-.99,.12)$ 
(unstable fixed point), $(0.41,-0.03)$ (stable fixed point), 
$(0.99,-0.12)$ (saddle), $(2.4,-0.2)$ (stable fixed point). 
(c) Enlarged image of the central portion in (b) showing the two fixed
points, one stable, and another a saddle.   In (b) and (c), the fixed points
$P_1,$ etc., may be identified near the respective numerical values.
}
\end{figure}
\begin{figure}[h]
%\centering\includegraphics[width=1\linewidth]{Figure3.eps}
\caption{Unstable regions in the $(\cos\theta, \omega)$ space shown by dark
regions, as obtained from the period map ${\bf M}$. Here, $\alpha=0.01$, 
$\kappa_{eff}=0.1$,
$q=0.1$. $(a)$ $\sin\theta_q=0$, 
$(b) \sin\theta_q=0.6, (c) \sin\theta_q=1$.}
\end{figure}
\begin{figure}[h]
%\centering\includegraphics[width=1.1\linewidth]{Figure4.eps}
\caption{Unstable regions corresponding to Figure $3$ in the $(h_{a\parallel}, 
h_{a\perp})$ space shown by dark regions, as obtained from the period map 
${\bf M}$. Here, $\alpha=0.01$, $\kappa_{eff}=0.1$, $q=0.1$. $(a)$ $\sin\theta_q=0$,
$(b) \sin\theta_q=0.6, (c) \sin\theta_q=1$. Not all points in the plane 
are accessible if one requires a homogeneous background, due to the condition 
$0\le h_{a\perp}^2\le \kappa_{eff}^2+\alpha^2\omega^2$.}
\end{figure}
\begin{figure}[h]
%\centering\includegraphics[width=1.1\linewidth]{Figure5.eps}
\caption{Unstable regions in the  $(\cos\theta, \omega)$ plane (left) and 
the corresponding regions in the $(h_{a\parallel}, h_{a\perp})$ plane (right). 
The parameter values correspond to the values for $(b)$ and $(c)$ in Figures
$3$ and $4$. Instabilities at $\omega_q=\omega$ are indicated by dots $(\cdot)$,
and those at $\omega_q=\omega/2$ by plus $(+)$. 
For $\sin\theta_q=0$, both $\rho$ and $\tilde{\rho}$ 
vanish as seen from equations $(\ref{critpara})-(\ref{critparb})$ and 
$(\ref{crit2para})-(\ref{crit2parb})$, and hence
no instability occurs. For $\sin\theta_q=1$, the instability at 
$\omega_q=\omega/2$ does not occur since $\tilde{\rho}$ vanishes as seen
from equations $(\ref{crit2para})-(\ref{crit2parb})$.}
\end{figure}

\end{document}